# Open Datasets for Satellite Radio Resource Control

*Husnain Shahid[1], Miguel Á. Vázquez[1], Musbah Shaat[1], Pol Henarejos[1], Nuria T Quijada[2]*
*Enrique Marinu[2]*

[1]*Centre Tecnològic de Telecomunicacions de Catalunya, Barcelona, Spain*
[2]*Hispasat, Madrid, Spain*
*\*hshahid@cttc.es*

**Keywords**: Open Datasets, Resource Management, Non-Terrestrial Networks, Artificial Intelligence

**Abstract**

In Non-Terrestrial Networks (NTN), achieving effective radio resource allocation across multi-satellite system, encompassing efficient channel and bandwidth allocation, effective beam management, power control and interference mitigation, poses significant challenges due to the varying satellite links and highly dynamic nature of user traffic. This calls for the development of an intelligent decision-making controller using Artificial Intelligence (AI) to efficiently manage resources in this complex environment. In this context, open datasets can play a crucial role in driving new advancement and facilitating research. Recognizing the significance, this paper aims to contribute the satellite communication research community by providing various open datasets that incorporate realistic traffic flow enabling a variety of uses cases. The primary objective of sharing these datasets is to facilitate the development and benchmarking of advanced resource management solutions, thereby improving the overall satellite communication systems. Furthermore, an application example focused on beam placement optimization via terminal clustering is provided. This assists in optimizing beam allocation task, enabling adaptive beamforming to effectively meet spatiotemporally varying user traffic demands and optimize resource utilization.

## 1. Introduction

In the context of NTN, the urge on efficiently utilizing the satellite resources has become increasingly essential and requiring attention. The emphasis is vitally important in order to enhance system performance while ensuring the reliable and cost-efficient services. The significance of these requirements become evident for the multi-orbital and multi-band satellite integrated networks. This is due to the fact of aggregating the muti-orbital satellite networks, impending with the complexity of varying satellite links combined with non-uniform fluctuations in user traffic demand across diverse use cases such as maritime, aeronautical, PPDR, just to mention a few [1]. Therefore, there is a need to design solutions to incorporate these complexities in order to ensure the effective performance in terms of radio resource management (RRM). In the context of RRM solutions, some notable contributions can be found to deal with these complexities by employing the adaptive plans and by leveraging the recently launched flexible payloads reconfiguration capabilities. As an example, with regard to flexible satellite payload capabilities concerning RRM, [2] studied the resource management problem and optimized the capacity management of the system by employing an objective function. While the authors in [3] provide insights into global resource management by incorporating the QoS metric and the channel conditions of users. Furthermore, a joint optimization tool is proposed in [4] which jointly optimize the beam width, power and bandwidth allocation to meet the variable traffic demand. Nevertheless, managing the number of resources at the larger scale, these methods show limited efficacy in addressing the complexities and thus leading to sub-optimal solutions. Responsively, AI-driven approaches have significant potential as a key contributor to the process of resource optimization. In this context, [5] introduces a deep reinforcement learning (DRL) algorithm, employing proximal policy optimization (PPO), to enhance optimization of unmet system demand and power consumption. Additionally, [6] addresses the challenge of inter-beam interference within multi-beam satellite systems. The is being achieved through intelligent user scheduling and strategic allocation of bandwidth and power resources, predominantly utilizing DRL methods. In another study, [7] delves into an optimal long term capacity allocation plan in a highly complex three-layer heterogeneous satellite network which uses the reinforcement learning as a learning framework to optimize the capacity allocation but in the realm of synthetic data.

It is widely acknowledged that AI paradigm heavily relies on datasets with large amount and diverse properties to learn the variable environment and provide efficient resource solutions. Therefore, this paper is solely focused on the provision of open AI-enabled datasets for potential use cases, tailored for resource management optimization task to gain more practical insights into the realistic and non-uniform traffic behaviour of the users. This will pave the way to the non-terrestrial communication systems research to develop the intelligent decision controllers based on AI. These controllers can leverage the realistic dataset properties and minimize the allocation of resources while serving the maximum users simultaneously.

Considering the objectives, the paper is organized in a following way: After introducing the significance of RRM, Section II focuses on maritime traffic dataset as a potential use case for NTN and discusses the attributes and instructions about utilization of this dataset. Section III presents aeronautical dataset providing insights about the flow of flights traffic while Section IV identifies the open datasets for Public Protection and Disaster Relief (PPDR) and residential traffic. Section V provides a practical example of utilization of one of these datasets for adaptive beamforming task which allows to manage the resources by serving the users' needs based on traffic demand. Finally, the conclusion is drawn in Section VI.

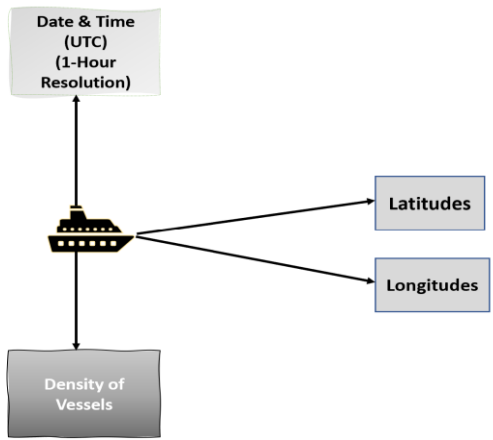

Figure 1. A general diagram of dataset attributes of marine traffic

## 2. Maritime Dataset

When it comes to satellite-enabled connectivity, NTN possess the ability to extend the coverage across expansive sea areas, effectively surpassing the limitations of terrestrial networks. Within this context, non-terrestrial networks emerge as having substantial potential to revolutionize the maritime sector. This potential translates into uninterrupted connectivity for passenger yachts, facilitating reliable communication among vessels, offshore platforms, maritime surveillance, and ultimately improving the efficiency of rescue operations while effectively meeting safety standards. Therefore, by enabling NTN, the maritime industry stands to gain immensely from enhanced connectivity and communication capabilities.

One of the significant datasets, stands out as a prime choice to be used to enhance the resource allocation performance within the maritime industry is obtained from a reliable data provider (VesselFinder & VT Explorer) [8]. This week-long dataset, capturing hourly data, has been acquired to provide comprehensive insights authentic marine traffic data derived from the Automatic Identification System (AIS) and covers the expansive Mediterranean Sea region. The maritime density map dataset provided herein is generated through MATLAB processing, utilizing the raw data acquired from VesselFinder. The dataset contains time information, geocoordinates (longitude, latitude) and normalized value of density flow as seen in Figure 1. The geocoordinates of a point where the information of vessels density is available are determined by grid-based method, where each grid spans an area of 100 km x 100 km using the haversine distance metric and the central point of each grid is a point where density of vessels is given. This idea to calculate the haversine distance is due to its capability to consider the Earth's surface curvature, leading to more precise calculations when compared to simpler methods like the Euclidean distance.

The objective to provide the tailored dataset here is to be employed for designing the RRM solutions to serve the maximum possible maritime users while taking into account the available satellites resources. This facilitates with the opportunity to indulge into optimizing the multiple resource allocation strategies, for instance, the provided data can be employed to train AI models that enable traffic forecasting using deep learning which offers the opportunity to pre-allocate the resources [9], resource scheduling to take full advantage of limited resources [10], and adaptive beamforming for specific users based on their throughput demand [4]. These trained models could serve users based on their traffic flow and throughput requirements at a given time instant. As a result, this could significantly contribute to organize the overall satellite system resources more efficiently.

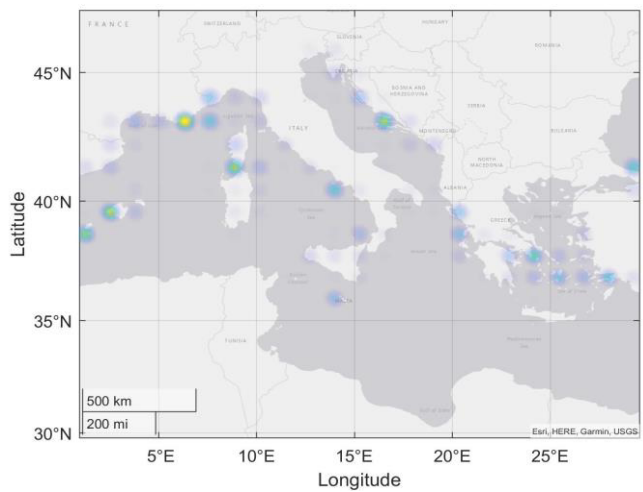

Figure 2. The maritime normalized traffic density map of an hour.

To establish this density map, the latitudes across the Mediterranean Sea are uniformly spaced using a factor of 0.9, equivalent to a haversine distance of 100 km with respect to latitudes, while maintaining a constant longitude. This procedure defines the vertical lines within a single grid. Subsequently, the longitudes of the Mediterranean Sea are uniformly spaced with a factor of 1.2, equivalent to a haversine distance of 100 km concerning longitudes, while keeping latitudes constant. This results in the formation of horizontal lines for the grid. By integrating these vertical and horizontal lines, an area of 100km is defined and each density value gives the density of traffic at every 100km region as shown in Figure 2. The offered dataset with this paper is in the form of comma-separated values (.CSV) and contain 3 columns where two columns represent the geocoordinates of central point within each grid cell and third row signifies the count of observed vessel. These provided numerical values facilitates to perform a thorough analysis and estimation of variations in maritime traffic across the dataset, yielding insights into the spatio-temporal distribution across diverse geographical regions within the Mediterranean Sea.

The given repository with this paper provides open access to the maritime dataset where each hour of the dataset file is stored in a .mat file format. The maritime traffic flow dataset (processed files) is accompanied by permissive license that allow users to access, commercially use, modify, and distribute the data freely. However, it is essential to give proper credits to the creators of the dataset (CTTC) as a sign of acknowledgement.

## 3. Aeronautical Dataset

Aeronautical is vitally important use case of NTN due to the unique challenges and requirements of aviation industry since the aeronautical operations cover diverse geographical areas across the globe including remote and oceanic regions. Therefore, it requires the NTN coverage where the terrestrial networks are unavailable, ensuring the seamless global coverage and connectivity. This is important because aviation operations entail continuous communication between aircraft, air traffic control and ground stations. Moreover, the requirements become critical in emergency situations when aircraft needs instant flight path adjustments due to intense weather conditions or in the risk of mid-air collision. Additionally, for the long-haul commercial flights, NTN enhances the passenger experience by offering seamless internet access and in-flight entertainments. Thereby, uninterrupted connectivity is essential for maintaining the safety operations since it enables the opportunity of global coverage, real time data transmission, air traffic management and passenger entertainment etc.

Where aeronautical operations necessitate the global coverage and seamless connectivity, the management of resources to serve the needs for such an immense air traffic become a critical concern as well. On top of that, dynamically changing traffic conditions poses challenging in managing the overall resources. Therefore, this RRM task is essential for maintaining the reliable communication links in terms of efficient spectrum allocation to enable interference free communication and to ensure that multiple aircrafts can communicate simultaneously without any signal degradation. Moreover, maintaining QoS, power allocation, beamforming and dealing with satellite handovers for highly dynamic nature of traffic is crucial. This can be handled using AI based algorithms if the information about the behaviour of air traffic is priori known. In order words, these algorithms can be trained by utilizing the features of air traffic pattern at any geographical region. These features could exhibit about how rapidly the air traffic changes for a given interval time and what is the density over a region. This information can be used to forecast the traffic and thereby schedule the resources to serve the needs of maximum users.

Regarding the traffic information, one of the promising air traffic datasets is given by the OpenSky Network with all the attributes that are typically require for RRM task and compatible to use as an input to train the AI- enabled radio resource intelligent controller to manage the resources given the realistic air traffic insights [11].

In order to retrieve the flight state information dataset, OpenSky Network uses automatic dependent surveillance-broadcast (ADS-B) and Mode S messages technology. The ADS-B works by having aircrafts determine their PVT (position, velocity and time) using satellite navigation or other sensor and broadcast to the entities for being tracked while each aircraft is identified with a unique address. As soon as the ADS-B message arrive at the OpenSky´s server, it establishes a record for the aircraft which is named "State Vector ". This state vector contains all the information that is required to track an aircraft for instance, the unique address, time information (Unix format), position, velocity, and heading etc. The typical extraction process of aircraft related state vector is shown in Figure 3. The available dataset spans from 25-05-2020 to 27-06-2022. The dataset covers full day and retrieves every Tuesday night for the preceding day. In other works, it provides a full snapshot of the Monday´s complete state vector data collected by OpenSky Network on Tuesday night only (a data of 24 hours in a week) with the updating rate one second [11]. This is actually useful and at least enough to understand the trend of the air traffic density and can be forecasted further using potential ML based algorithms.

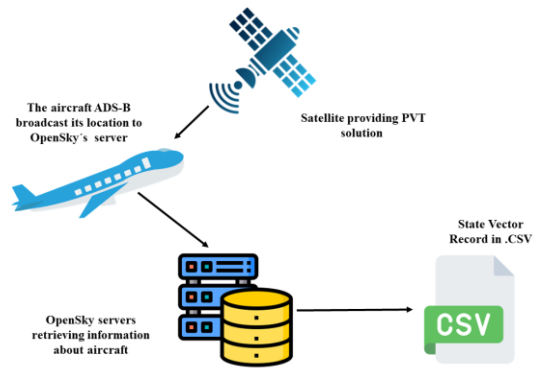

Figure 3. The overall scheme of aircraft data acquisition by OpenSky Networks [11]

## 4. PPDR and Residential Dataset

PPDR and Residential traffic are the essential use cases among the prime candidates that to be facilitated with NTN. The purpose of considering the PPDR and residential traffic for NTN can be partitioned into two events. The first is related to the temporary events that appear in case of emergency, for instance due to the natural disasters, terrestrial network may be severely impaired. The provision of NTN services enables the relief for the affected individuals to seek emergency assistance and communicate about their situation to authorities. Similarly, it empowers the authorities in the affected areas to effectively communicate and fasten the rescue operations in the absence of terrestrial facilities. The second event corresponds to the permanent events, when there is a need for the provision of connections in remote villages that lack the overall terrestrial infrastructure. In such scenario, NTN enables users to access the internet, make voice calls and exchange text messages.

However, acquiring the dataset intended for the PPDR and residential use case can pose challenges, primarily due to the complexities in identifying regions characterized by inadequate terrestrial communication infrastructure so that resources allocation from NTN can be restricted to only those areas, since in such areas network performance might fall short than the required. Additionally, strict obligations on personal data privacy and security laws further make the utilization more challenging. These laws enforce rigorous protocols for the collection, storage, and publication of any data associated with individuals within the EU, including their mobility and geolocalization. Compliance with regulations like the general data protection regulation (GDPR) or specific national guidelines, such as those issued by the CNIL in France, is of utmost importance in this regard.

Fortunately, Ookla's open data initiative provides a reliable repository that grants access to adaptable data for resource allocation, allowing to modify in accordance with specific goals [12]. Moreover, the Urban Data Platform Plus (UDPP), provided by the European Commission (EC), further processes the Ookla's Speedtest data and provides valuable insights into both fixed and mobile networks performance, segmented at the municipality level across the European region [13], thereby simplifying the work process at regional level. In addition, the number of people residing in an area is helpful information to manage satellite resources in order to offer the sufficient throughput capacity to all users. Relevant to this, Meta provides the open high-resolution population density map involves employing state of the art computer vision techniques [14]. Although, it does not exactly provide the number of active user terminals but taking into account the information of people quantity, an estimated number of user terminals corresponding to a percentage of population can be enabled to serve by managing the throughput capacity accordingly.

Regarding the creation description of these datasets, Ookla´s dataset offers the global fixed broadband and mobile network performance in the form of tiles. The tiles comprise into zoom level 16 Web Mercator projection, measuring approximately 610.8 by 610.8 meters at the equator. This zoom level has a meaning of covering the earth region, for instance, zoom level 0 gives the size of tile equivalent to entire globe while zoom level 1 represents into 4 tiles The fixed broadband is about the measurements taken from mobile devices with non-cellular type connection i.e., Wi-Fi or Ethernet whereas the mobile network measurements are taken from mobile devices with cellular type connection, for instance, 4G LTE and 5G NR. The network performance statistics in Ookla dataset are acquired by using the mobile devices performing speed tests. The tests are performed at the random hours with several devices, or a single device performs multiple tests in some scenarios. The measurements given by all devices acquired from each test are then aggregated to get mean value in each tile. The provided key performance indicator (KPIs) in dataset such as average latency and average speed for both upload and download are based on devices participation and tests repetitions. Therefore, higher the number of device participations and tests lead to more accurate values of KPIs.

The dataset spans from the period of Q1 2019 to the recently complete quarter Q2 2023 and being updated every quarter year (three months) [12]. Hence, a continuous updated information can be acquired after a defined interval of time.

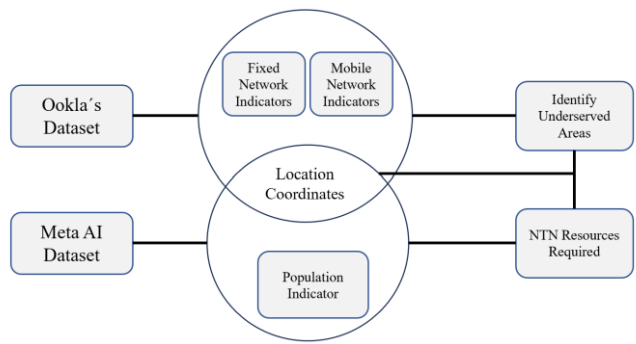

Figure 4. The integrated data guidelines to identify the resources required for underserved/remote areas.

On the other hand, the generation of high-resolution population density dataset involves applying state of the art computer vision techniques on high resolution (50 cm per pixel) satellite imagery data integrated with the publicly available census data. The concept underlying the generation of statistics for population density involves training the convolutional neural network (CNN) to identify the buildings from public accessible mapping service such as OpenStreetMap (OSM) to count the statistical number of people. The population density map exhibits impressive accuracy level, providing data with 30 x 30-meter grid resolution, making it reliable and authentic dataset for utilization [14].

The idea is to identify the areas where fixed or mobile network performance is below the minimum value typically required for communication. This information is useful to fil the gaps by NTN to better serve the user needs in during both temporary and permanent events. Subsequently, the population density map gives insights into the statistical number of people residing in any region. Hence, the aggregation of both these datasets may assist in defining the NTN resources in those identified areas where the network performance is not up to the mark. The insufficient network performance may be due to network congestion, but the educated guess is, lower the bit rate, higher the probability of employing NTN services. The general aggregation scheme is shown in Figure 4.

## 5. Adaptive Beamforming Example

Taking the maritime dataset into account, an adaptive beamforming example is presented here which incorporates the information of maritime traffic flow and perform flexible beamforming accordingly. This further optimizes the resource allocation by projecting the beams only at those geocoordinates where the marine traffic is present. In this essence, the first step is to initialize the clusters which makes the placement of beams easier due to prior information of marine traffic. The cluster centers are also be referred as beam centers to serve over the region. The following step is then allocating the random data demand to each vessel which they will be requested. The demand varies from 1Mb/s to

20Mb/s for necessities such as messages, audio calls etc. To further optimize the beams placement based on the data demand of users, the clusters are relocated in such a way that the vessels with higher demand would be closer to the cluster or beam centers and able to get themselves satisfied with the requested demand since being closer to the beam center would experience high signal to noise ratio. This is being done by using the weighted K-means clustering algorithm which is an un-supervised machine learning technique. Going into deeper insights of the weighted K-means algorithm, each beam determines the vessels that have minimum distance to that beam and takes the weighted average of demands over all vessels in that beam. It will ultimately decide the position of central point of beam. This process will be repeated for number of iterations that are predefined which is chosen to be 1000 in our case. The beam centers coordinate as clusters and the vessels belong to each cluster can be seen in Figure 5.

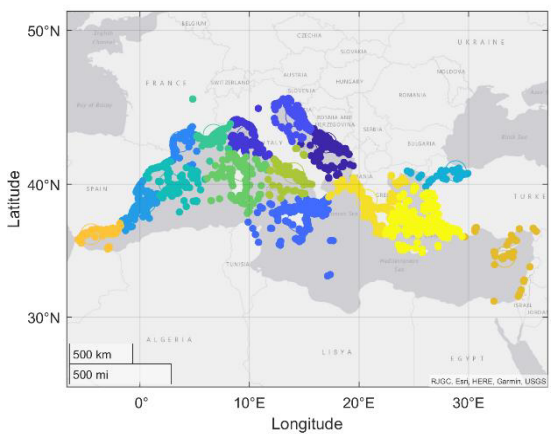

Figure 5. The same color vessels belong to same cluster and transparent big circles are beam/cluster centers.

This demand based adaptive beamforming assists in optimizing the resource management which can be analysed by assessing the demand distribution in each beam. In this context, two beamforming systems are studied 1) Fixed beams 2) Adaptive beams. The comparison of demand distribution in both systems depicted in Figure 6., reveals that, for the fixed beams scenario, the demand request is very unequal since the probability of having zero demand is higher which means that no user is there to request any demand or no user is present in that geographical region, but the beams are being served at that location, hence if there is no demand, the capacity will be unused.

The same probability of having zero demand is lower or even zero for adaptive beam scenario, meaning that all the beams are serving the users and no beam is empty. On the other hand, the fixed beams cumulative density function (CDF) reaches at highest level in a quick manner than the adaptive beam system. It means, most of the beams in fixed beam system are serving the higher density of users so if there are large number of users with higher demand and it´s most probable that the capacity requirements will be unmet in some beams and leading to inefficient resource allocation. While for adaptive beamforming based on user traffic information, the CDF is increasing in almost a uniform way revealing that each beam is serving the uniform number of users at a given time, making the system more efficient in terms of RRM.

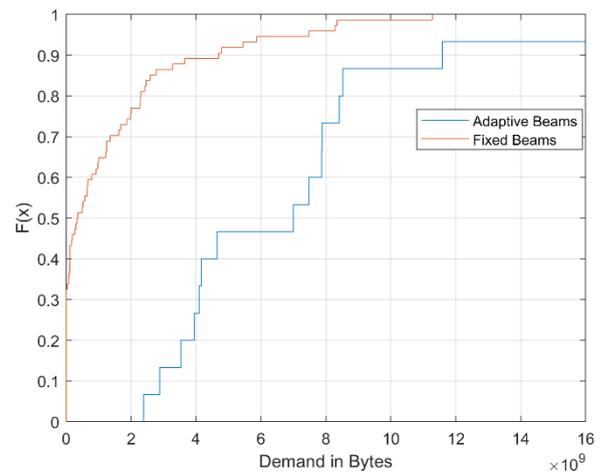

Figure 6. The CDF comparison for both Adaptive and Fixed Beams system

Another KPI is Jain´s Fairness Index [15], which gives a single number about how evenly the demand is distributed among all the beams. The equation to calculate the JF index is given as,

$$JF = \frac{(\sum_{k=1}^{K} D_k)^2}{K \sum_{k=1}^{K} D_k^2} \qquad (1)$$

Where $D_k$ is the summation of demands of all users in beam $K$. The JF values ranges between (1/number of beams and 1) where 1 means most efficient. The calculated JF index is **0.76-0.80** for adaptive beams while **0.27** for fixed beams system concluding the efficiency of system when prior information of traffic is known.

## 4 Conclusion

The urge of radio resource management within NTN holds paramount importance, as it directly impacts the overall communication system capacity. However, this task in a heterogeneous environment poses challenges due to the dynamic nature user traffic demands. In order to cope with these challenges, it is imperative to consolidate these variabilities in designing AI based RRM solutions. Therefore, a prior information about these variabilities can assist the algorithms to converge and effectively manage the satellite resources. In this regard, this document provides an opportunity to identify and the provision of open datasets as potential candidates for AI- enable RRM framework for multiple use case such as maritime, aeronautical, and PPDR etc.

Regarding the open datasets, this document offers the explicit access and overview of their relevance with each defined use case. Moreover, it gives valuable insights into how these datasets are compatible for being used as an input to AI based algorithms and thereby beneficial for managing the resources given the realistic numerical values of traffic demand belongs to various services. Ultimately, an adaptive beamforming example is provided using un-supervised machine learning algorithm using realistic maritime dataset which give insights about how the available dataset optimize the resources.

## 5 Acknowledgements

This work was supported by the HORIZON-CL4-2021-SPACE-01 project "5G+ evoluTion to mutioRbitAl multibaNd neTwORks" (TRANTOR) No. 101081983 and 6G-NTN project funded by smart networks and services joint undertaking (SNS JU) under the European Union´s Horizon Research and Innovation Program under grant agreement No. 101096479 while a part of this work is also supported by the R+D+i project (PID2020-115323RB-C31) funded by MCIN/AEI/ 10.13039/501100011033.

## Data Availability Statement

The datasets are openly accessible through the following links https://datasets.cttc.es/ and https://cloud.cttc.es/index.php/s/Z9F4gSnRsxFG3BM

## 6 References

[1] Al-Hraishawi, H., Lagunas, E. and Chatzinotas, S. Traffic simulator for multibeam satellite communication systems. 10th Advanced Satellite Multimedia Systems Conference and the 16th Signal Processing for Space Communications Workshop (ASMS/SPSC), 2020, pp. 1-8.

[2] Cocco, G., De Cola, T., Angelone, M., Katona, Z. and Erl, S. Radio resource management optimization of flexible satellite payloads for DVB-S2 systems. IEEE Transactions on Broadcasting, 64(2), 2017, pp.266-280.

[3] Su, Y., Liu, Y., Zhou, Y., Yuan, J., Cao, H. and Shi, J.. Broadband LEO satellite communications: Architectures and key technologies. IEEE Wireless Communications, 26(2), 2019, pp.55-61.

[4] Honnaiah, P.J., Maturo, N., Chatzinotas, S., Kisseleff, S. and Krause, J.. Demand-based adaptive multi-beam pattern and footprint planning for high throughput GEO satellite systems. IEEE Open Journal of the Communications Society, 2021, pp.1526-1540.

[5] Luis, J.J.G., Guerster, M., del Portillo, I., Crawley, E. and Cameron, B., Deep reinforcement learning for continuous power allocation in flexible high throughput satellites. In IEEE cognitive communications for aerospace applications workshop 2019, pp. 1-4.

[6] Leng, T., Wang, Y., Hu, D., Cui, G. and Wang, W.. User-level scheduling and resource allocation for multi-beam satellite systems with full frequency reuse. China Communications, 2022, pp.179-192.

[7] Jiang, C. and Zhu, X. Reinforcement learning based capacity management in multi-layer satellite networks. IEEE Transactions on Wireless Communications, 19(7), 2020, pp.4685-4699.

[8] The FREE AIS vessel tracking platform (https://www.vesselfinder.com/).

[9] Yao, H., Tang, X., Wei, H., Zheng, G. and Li, Z. Revisiting spatial-temporal similarity: A deep learning framework for traffic prediction. In Proceedings of the AAAI conference on artificial intelligence 33(01), pp. 5668-5675.

[10] Fourati, F. and Alouini, M.S.. Artificial intelligence for satellite communication: A review. Intelligent and Converged Networks, 2(3),2021, pp.213-243.

[11] The accessible platform of air traffic dataset (https://opensky-network.org/datasets/states/).

[12] Speedtest by Ookla Global Fixed and Mobile Network Performance Maps was accessed from (https://registry.opendata.aws/speedtest-global-performance).

[13] The European Commission Urban Data Platform Plus (https://urban.jrc.ec.europa.eu/).

[14] The country wise population density dataset using AI, provided by META (https://dataforgood.facebook.com/dfg/docs/methodology-high-resolution-population-density-maps).

[15] Jain, R.K., Chiu, D.M.W. and Hawe, W.R A quantitative measure of fairness and discrimination. Eastern Research Laboratory, Digital Equipment Corporation, Hudson, MA, 1985, pp-21.